\documentclass[12pt, prd, showpacs]{revtex4}
\input{tcilatex}

\begin{document}

\title{Ultimate gravitational mass defect }
\author{O. B. Zaslavskii}
\affiliation{Department of Mechanics and Mathematics, Kharkov V.N.Karazin National
University, \\
Svoboda Square 4, Kharkov 61077, Ukraine}
\email{ozaslav@kharkov.ua}

\begin{abstract}
We present a new type of gravitational mass defect in which an infinite
amount of matter may be bounded in a zero ADM mass. This interpolates
between effects typical of closed worlds and T-spheres. We consider the
Tolman model of dust distribution and show that this phenomenon reveals
itself for a solution that has no origin on one side but is closed on the
other side. The second class of examples corresponds to smooth gluing
T-spheres to the portion of the Friedmann-Robertson-Walker solution. The
procedure is generalized to combinations of smoothly connected T-spheres,
FRW and Schwarzschild metrics. In particular, in this approach a finite
T-sphere is obtained that looks for observers in two R-regions as the
Schwarzschild metric with two different masses one of which may vanish.
\end{abstract}

\pacs{04.70.Bw, 04.20 Gz, 04. 20 Jb.}
\maketitle

% It is always \today, today, but any date may be explicitly specified


\section{Introduction}

One of the remarkable features of general relativity is a gravitational mass
defect because of which a total ADM mass $m$ measured of an external
observer at infinity may significantly differ from the proper mass $M$ -
total amount of matter contained in the system, $m<M$ \cite{lan}. There
exist two distinct situations of extremely strong gravitation binding. (i)
It can arise in a semi-closed world connected to the external region via
"throat". When this throat shrinks, a completely closed world appears
separated from the external region, its mass $m$ tending to zero (see
monographs \cite{har} - \cite{nf} and references therein). (ii) Another type
of the gravitation mass defect is inherent to systems without the centre of
symmetry that are able to accumulate an infinite amount of matter but reveal
themselves for an external observer as a body of a finite mass $m$ \cite%
{nov2}. In particular, this is the property typical of so-called T-spheres 
\cite{rub1} - \cite{rub3} (see below) which in the case of the dust source
generalize the Kantowski - Sachs class \cite{kant} of exact solutions. The
mass $m$ for such models is constant but the T-region can extend infinitely
long. (We adhere the classification and terminology introduced in \cite{nov1}%
: if the gradient of the areal radius $R$ is space-like, it is called
R-region, if it is time-like, it is called T-region. In doing so it is
implied that this gradient does not identically vanish.) Being glued to a
vacuum region, such a system reveals itself in the outer space as a
"T-sphere" of a finite mass. Thus, in the case (i) we are faced with a 
\textit{finite} $M$ and \textit{zero }$m$, in the case (ii) $M$ is \textit{%
infinite}, $m$ is \textit{finite}. It was specially stressed in \cite{rub3}
that the nature of the phenomenon under discussion in these two situations
is qualitatively different. The basic aim of the present note is to point
out that, nevertheless, both cases can be combined in such a way that they
lead to a new type of the gravitational mass defect which is the most
possible strong one: an \textit{infinite} $M$ corresponds to a \textit{zero} 
$m$.

\section{Preliminaries: Tolman model and definition of mass}

For simplicity, we restrict ourselves by perfect dust that admits exact
solutions \cite{tol}. We begin with short description of their properties.
As for dust there exists the frame which is simultaneously synchronous and
comoving, the metric under the assumption of spherical symmetry can be cast
into the form%
\begin{equation}
ds^{2}=-d\tau ^{2}+b^{2}(\chi ,\tau )d\chi ^{2}+R^{2}(\chi ,\tau )d\omega
^{2}\text{, }d\omega ^{2}=d\theta ^{2}+d\phi ^{2}\sin ^{2}\theta \text{.}
\label{tbl}
\end{equation}

Usually, it is implied that the derivative $R^{\prime }$ does not vanish
identically (hereafter, prime and dot denote derivatives with respect to $%
\chi $ and $t$, correspondingly). Then the model admits known Lema\^{\i}%
tre-Tolman-Bondi (LBT) family of solutions which describe an inhomogeneous
collapse of dust (or its time reversal). However, there is also one more
brunch that arises as a special solution of Einstein equations with the
areal radius $R=R(t)$ not depending on a spatial coordinate \cite{rub1}, 
\cite{rub2}, \cite{datt}. The special solution under discussion (called
"T-spheres" or "T-models" in \cite{rub1}, \cite{rub2}) possesses a number of
unusual properties. Thus, there are two qualitatively different branches of
the metric of the type (\ref{tbl}).

\subsection{LBT solutions}

The LBT solutions can be represented in the form

\begin{equation}
b^{2}=\frac{R^{\prime 2}}{1+f(\chi )}\text{,}  \label{11}
\end{equation}%
where $f(\chi )$ should satisfy the inequality $f+1\geq 0$, otherwise being
arbitrary. The time evolution is governed by equation%
\begin{equation}
\dot{R}^{2}=\frac{F(\chi )}{R}+f\text{,}  \label{time}
\end{equation}%
the energy density%
\begin{equation}
8\pi \rho =\frac{F^{\prime }}{R^{\prime }R^{2}}\text{,}
\end{equation}%
where $F(\chi )$ is another function parametrizing the solution, $F(\chi
)=2m(\chi )$ (we assume the system of units in which $c=G=1$), the quantity $%
m(\chi )$ is an effective mass introduced first by Lema\^{\i}tre \cite{lem}
and later discussed in \cite{mis} - \cite{int}. For a generic
spherically-symmetrical spacetime it is defined according to%
\begin{equation}
m(\chi )=\frac{R}{2}[1-\left( \nabla R\right) ^{2}]\text{.}  \label{mass}
\end{equation}%
Its significance consists in that it obeys the simple equation that follows
from Einstein equations:%
\begin{equation}
m_{,a}=4\pi R^{2}(T_{a}^{b}-T\delta _{a}^{b})R_{,b}\text{,}  \label{m'}
\end{equation}%
where $a=0,\chi $ and $T=T_{a}^{a}$, $T_{\mu }^{\nu }$ is the stress-energy
tensor. In the case of dust its only non-vanishing component in the
co-moving frame $T_{0}^{0}=-\rho $. It is easy to check that for LTB
solutions (\ref{11}), (\ref{time}) indeed $m(\chi )=\frac{F(\chi )}{2}$. Eq.
(\ref{m'}) with $a=0$ gives in this case $\dot{m}=0$, so that $m(\chi )$ is
the integral of motion.

If a configuration of matter (dust in our case) is glued to the vacuum at
some $\chi _{0}$, the condition of smoothness requires the continuity of the
mass, so that $m(\chi _{0})=m$, where $m$ is the ADM mass of the
corresponding Schwarzschild solution in the vacuum region \cite{mis}.

The proper mass is defined according to%
\begin{equation}
M(\chi )=4\pi \int dlR^{2}\rho =4\pi \int d\chi bR^{2}\rho \text{,}
\label{M}
\end{equation}%
where $dl$ is the proper distance. One can easy to check that $M(\chi )$ is
also an integral of motion in the Tolman model.

The function $f=f(\chi )$ can have any sign. For our purposes, as will be
seen from what follows, it is sufficient to restrict ourselves to the case $%
f<0$ (so-called elliptic case). Then

\begin{equation}
R=\frac{F}{2(-f)}(1-\cos \eta )\text{,}  \label{R}
\end{equation}%
\begin{equation}
\eta -\sin \eta =\frac{2(-f)^{3/2}}{F}(\tau -\tau _{0})\text{, }\tau
_{0}=\tau _{0}(\chi )\text{,}  \label{t}
\end{equation}%
$\tau _{0}$ is the time of a big bang (if $\tau \geq \tau _{0}$) or\ big
crunch (if $\tau \leq \tau _{0}$).

\subsection{T-spheres}

If $R=R(\tau )$, it is found in \cite{rub1}, \cite{rub2} that

\begin{equation}
R=\frac{R_{0}}{2}(1-\cos \eta )\text{, }\tau =\frac{R_{0}}{2}(\eta -\sin
\eta )\text{, }  \label{t1}
\end{equation}

\begin{equation}
b=\varepsilon \cot \frac{\eta }{2}+2M^{\prime }(1-\frac{\eta }{2}\cot \frac{%
\eta }{2})\text{, }\varepsilon =0\text{, }\pm 1\text{,}  \label{t2}
\end{equation}%
where $R_{0}=const$. Such a kind of solutions cannot be considered as a
particular case of LBT family but, rather, represents a separate branch
(nonetheless, T-models can be obtained from a special class of LBT solutions
due to a special limiting transition \cite{tt}). One can easily see from (%
\ref{mass}) that now the quantity $m=\frac{R_{0}}{2}$ is a constant not
depending on $\chi $. For a fixed moment of time the spatial geometry
represents a hypercylinder.

One more distinctive feature of T-models consists in that they do not have a
centre of symmetry and, thus, represent an ideal tool for linking different
universes (see, for example, short discussion in recent paper \cite{gair}),
in this sense being a natural generalization (and materialization) of the
Einstein-Rosen bridge.

\section{LBT solutions with one origin}

Now we demonstrate the example of LTB solutions such that the geometry from
the left side approaches that of an infinitely long hypercylinder while from
the right side it borders with vacuum and in some limit closes. We will see
now that this ensures the ultimate gravitational effect under discussion.
Let $f$ near some $\chi =\chi _{0}$ have the asymptotic behavior

\begin{equation}
f=-1+k(\chi -\chi _{0})^{2}\text{, }k>0\text{. }  \label{fk}
\end{equation}%
We also take $F^{\prime }$ to be finite at $\chi _{0}$. Then the proper
distance between points with $\chi =\chi _{0}$ and $\chi \neq \chi _{0}$%
\begin{equation}
l=\int d\chi \frac{\left\vert R^{\prime }\right\vert }{\sqrt{1+f}}
\end{equation}%
diverges like $l=-\frac{\left\vert R^{\prime }(\chi _{0})\right\vert }{\sqrt{%
k}}\ln (\chi -\chi _{0})$ and $\frac{\partial R}{\partial l}=R^{\prime }%
\frac{\partial \chi }{\partial l}\sim \chi -\chi _{0}\rightarrow 0$. Thus,
the coordinate $\chi $ becomes degenerate in this limit, the geometry of
manifold tends to that of an infinite hypercylinder, $R=const=R(\chi _{0})$.
In doing so, the density $\rho \rightarrow \rho (\chi _{0})=const$, so that
according to (\ref{M}) $M$ diverges.

We are still left with the freedom to choose the behavior of functions $%
f(\chi )$ and $F(\chi )$ in the region $\chi >\chi _{0}$. Let us choose $%
F(\chi )$ to be monotonically decreasing and vanishing at some $\chi _{1}$.
It means that the mass function $m(\chi )$ also vanishes and the world
becomes closed from the point of an external observer, its Schwarzschild
mass $m=0$. In general, the Tolman model can contain singularities such as
surface layers or shell crossing \cite{cross}, \cite{dis}, so that we should
check that with our choice these singularities do not appear. It follows
from the results of the aforementioned works that one may achieve a model
with neither surface layer nor shell crossing demanding $F^{\prime }<0$, $%
\tau _{0}=0$, provided $\frac{F^{\prime }}{F}-\frac{3}{2}\frac{f^{\prime }}{f%
}<0$ in the region $\chi _{0}\leq \chi \leq \chi _{1}.$Then $R$ is
monotonically decreasing function of $\chi $.

Thus, the above example supplies us with the configuration which is
everywhere regular (until a moment of a big crunch or after big bang), the
proper mass $M$ is infinite, the mass $m=0$. The geometry of the solution
under discussion is such that it has no origin on the left side but has an
origin ($R=0$) on the right side. One can also consider such a profile of
the function $m(\chi )$ that at some $\chi _{0}$ it becomes constant: $%
m=m(\chi _{0})>0$ for $\chi \geq \chi _{0}$. Then we will have a semi-closed
world that in the limit $m\rightarrow 0$ turns into a closed one.$\,$

It is worth noting that in our case inequalities between $m$ and $M$ derived
in \cite{ed} for any bounded spherically symmetrical distribution of matter
do not apply since our system is not bounded and has an infinite proper
volume.

\section{Piecewise constructions}

We will consider now piecewise constructions obtained by gluing between some
manifolds corresponding to explicit exact solutions. The general
classification of types of smooth gluing between different pieces of
solutions in the Tolman model was elaborated in \cite{tbl} but for our
purposes it is sufficient to take the simplest examples. Let the matter be
described by the Friedmann-Robertson-Walker (FRW) solution in the inner
region $\chi \leq \chi _{0}$ with $\chi _{0}>\frac{\pi }{2}$. Outside dust
the spacetime is vacuum and is described by the Schwarzschild (S) metric.
The FRW and S solutions are easily glued at the boundary surface $\chi =\chi
_{0}$ (see \cite{zn} for details and references). In doing so, $\tau -\tau
_{0}=\int_{R}^{R_{0}}dr(\frac{r_{g}}{r}-\frac{r_{g}}{R_{0}})^{-1/2}$, $\tau
_{0}$ and $R_{0}$ are some constants, $r_{g}=2m\ $is the gravitational
radius, $m$ is the mass measured by an observer at infinity. Inside the
matter (see, for example, the problem after Sec. 103 in \cite{lan}) the
metric takes the form (\ref{tbl}) with 
\begin{equation}
b=a(\tau ),R=a(\tau )\sin \chi \text{,}  \label{ba}
\end{equation}
where%
\begin{equation}
a=\frac{r_{0}}{2}(1-\cos \eta )\text{, }\tau =\frac{r_{0}}{2}(\eta -\sin
\eta )\text{, }  \label{at}
\end{equation}%
$r_{0}$ being the constant, $\chi _{\ast }\leq \chi \leq \chi _{0}$. For any 
$\chi _{0}$ the quantity $R(\tau ,\chi _{0})$ coincides with $R$ calculated
according to the outer S metric.

Let the region $\chi \leq \chi _{\ast }$ correspond to the T-sphere. Thus,
the whole sequence of spacetimes (from the left to the right) can be written
as T-FRW-S$.$ Gluing T-sphere to the metric (\ref{tbl}) along the surface $%
\chi =\chi _{\ast }$ produces the extrinsic curvature tensor $K_{\mu }^{\nu
} $ whose only non-zero component $K_{2}^{2}=-\frac{R^{\prime }}{Rb}$. If $%
\chi _{\ast }$ is such that $R^{\prime }=0$, $R\neq 0$ in the portion of the
metric that borders with the T-region, $K_{\mu }^{\nu }$ turns out to be
continuous across this surface and, hence, according to general criterion 
\cite{k} gluing is smooth. Therefore, we may choose the surface $\chi _{\ast
}=\frac{\pi }{2}$ that satisfies this criterion and match the FRW metric to
that of the T-model. In doing so, it follows from (\ref{t1}), (\ref{at}),
and (\ref{ba}) that $R$ calculated from both sides (described by the FRW
metric and that of the T-sphere) is the same, if we identify the constants $%
R_{0}=r_{0}$ entering both branches of solutions (LBT ones and T-spheres).

Now we may exploit the facts that (a) smooth gluing demands coincidence of
effective masses $m(\chi )$ \cite{mis} and (b) the mass $m$ does not depend
on $\chi $ for T-models \cite{rub1}. As a result, for $\chi \leq \frac{\pi }{%
2}$ $m=const=m(\frac{\pi }{2})$. But for FRW solutions the mass $m(\chi )=%
\frac{r_{0}}{2}\sin ^{3}\chi $, as it follows from (\ref{mass}), (\ref{ba}),
(\ref{at}). Substituting $\chi =\frac{\pi }{2}$, we obtain that $m=\frac{%
r_{0}}{2}$. The outer observer in the S asymtotically flat region measures
the mass $m(\chi _{0})$. In the limiting case $\chi _{0}\rightarrow \pi $
this mass $m(\chi _{0})\rightarrow 0$. Meanwhile, as now $\chi $ is not
bounded from below and extends infinitely to the region $\chi \leq \frac{\pi 
}{2}$ corresponding to the T-sphere, the total proper mass $M$ (amount of
matter) diverges that is typical of T-models \cite{rub1} - \cite{rub3}.
Thus, gravitational binding is absolute in the sense that an infinite $M$
corresponds to a vanishing $m$.

In the example we have just discussed the T-region extends to minus
infinity. However, we may abrupt it and glue smoothly to the left S metric
at some $\chi _{1}$ in the same manner to obtain the model of the type
S-T-FRW-S. Repeating all steps we see now that the left observer in its
R-region (call it R$_{2}$) of the S geometry measures the mass $m(\chi _{1})=%
\frac{r_{0}}{2}$, whereas the right one in its R-region (call it R$_{1}$)
measures $m(\chi _{0})$. In the limit $\chi _{0}=\pi $ the mass $m(\chi
_{0})\rightarrow 0$ but we want to stress that, in contrast to the case
considered in \cite{zn}, now the left world separated from the right one is 
\textit{not} \textit{closed. }Moreover, the external left observer measures
its nonzero mass. Usually, two Schwarzschildian R-regions do not appear at
all when one sews the solution in vacuum with that in matter \cite{zn1} that
occupies some region with $0\leq \chi \leq \chi _{0}$, with $\chi =0$
corresponding to the centre where $R=0$. However, now this argument does now
work since T-region does not have a centre of symmetry at all. It was
pointed out in literature \cite{dense} that Tolman model admits
generalization of the Kruskal construction to the case of non-zero density.
The spacetime under discussion which is "made" from known solutions and in
which dust is bound between pure vacuum regions, represents another example
of such a kind.

Usually, the property $m=0$ and close character of the geometry are
considered to be tightly connected and, in quantum theory, this opens the
possibility of quantum creation of such worlds \cite{t} - \cite{vilcr}.
However, we saw that in our case the world with a vanishing mass is open
from one side and possess an infinite proper volume that seems to be an
obstacle to its quantum creation.

\section{Conclusions}

Within Tolman model, we have manage to find explicit solutions which (i)
contain no surface layers and shell crossing, (ii) posses an infinite proper
mass and (iii) zero ADM mass. Two kinds of such solutions has been found.
The first one is based on LBT solutions only, the second one represents some
combinations of LBT solutions and T-spheres. These two cases share the
common property: the geometry is essentially different on two sides. On one
side, it approaches an infinitely long hypercylinder in LBT case or
represents such a hypercylinder precisely in the T-sphere case. On the other
side, the geometry is similar to that of closed world like in known FRW
solutions. In passing, we showed that constructions based on gluing between
T-spheres and other Tolman solutions (including pure vacuum ones) lead to
asymmetric constructions like two asymptotically flat regions with different
masses. In this sense, dust in such solutions represents materialization of
the Einstein-Rosen bridge between different universes.

\end{document}